\documentclass[twocolumn,preprintnumbers,amsmath,amssymb]{revtex4}
\usepackage{graphicx,epsfig}
\usepackage{dcolumn}
\usepackage{bm}
\begin{document}

\title {Dynamical Heterogeneity and the interplay between activated and 
mode coupling dynamics in supercooled 
liquids 
}
\author{ Sarika Maitra Bhattacharyya$^{\dagger}\footnote
{Electronic mail~:sarika@sscu.iisc.ernet.in}$, 
Biman Bagchi$^{\dagger}$\footnote{Electronic mail~:bbagchi@sscu.iisc.ernet.in} \\
and \\
 Peter G. Wolynes$^{\ddagger}$\footnote
{Electronic mail~:pwolynes@chem.ucsd.edu}}

\affiliation{$^{\dagger}$ Solid State and Structural Chemistry Unit, 
Indian Institute of Science, Bangalore 560 012, India.\\
$^{\ddagger}$Department of Chemistry and Biochemistry, 
University of California at San 
Diego, La Jolla, California 92093-0371}

\begin{abstract}
We present a theoretical analysis of the dynamic structure factor (DSF)
of a liquid at
and below the mode coupling critical temperature $T_c$, by developing
a self-consistent theoretical treatment which includes the contributions both from
continuous diffusion, described using general two coupling parameter ($F_{12}$)
mode coupling theory (MCT), and  from the
activated hopping,  described using the random first order transition (RFOT) 
theory, incorporating the effect of dynamical heterogeneity. 
The theory is valid over the whole temperature plane and shows 
correct limiting MCT like behavior above $T_{c}$ and goes over to the RFOT 
theory near the glass transition temperature, $T_{g}$. Between 
$T_{c}$ and $T_{g}$, the theory predicts that neither the continuous diffusion, 
described by pure mode coupling theory, 
nor the hopping motion alone suffices
but both contribute to the dynamics while interacting 
with each other.
We show that the interplay between the two contributions conspires to modify the
relaxation behavior of the DSF from what would be predicted by a theory with 
a complete static Gaussian barrier distribution in a manner that may be 
described as a facilitation effect.
Close to $T_c$, coupling between the short 
time part of MCT dynamics and hopping reduces the stretching given by the F$_{12}$-MCT
theory significantly and
accelerates structural relaxation. As the temperature is progressively lowered
below $T_c$, the equations yield a crossover from MCT dominated regime to the
hopping dominated regime. 
In the combined theory the dynamical heterogeneity is modified because the
low barrier components interact with the MCT dynamics to enhance the relaxation rate
below $T_c$ and reduces the stretching that would otherwise arise from an 
input static 
barrier height distribution. Many of these results can be explained 
from an analytical treatment of the combined
equation of motion.
\end{abstract}
\maketitle
\section{Introduction}

In an earlier article we showed how to connect self-consistently the mode
coupling theory (MCT) with the random first order transition theory (RFOT) 
to describe the dynamics of a liquid above and below the mode coupling 
transition temperature, $T_{c}$ \cite{sbp}. 
The resulting dynamics 
includes both the diffusive dynamics described by MCT and the hopping dynamics 
described by RFOT theory. Although other earlier attempts to include 
both hopping and mode coupling dynamics within one theoretical
scheme have been made \cite{gotze,das}, the merit of our calculation was
the use of
hopping dynamics, determined  
via the 
RFOT theory, thus acknowledging in accord with experiments  
that the hopping rate decreases with the configurational entropy. Because of 
the feedback between the structural 
relaxation (which includes contribution from both continuous and hopping 
dynamics) and the viscosity, hopping has a non-linear effect on the total
dynamics. Due to the self-consistent nature of the 
calculation there was hopping induced softening of the growth of the frequency 
dependent viscosity with decreasing temperature 
and this in turn helped the relaxation of the MCT 
contribution to the structural relaxation. Thus, the theory 
predicts that below $T_{c}$ along with the input hopping dynamics 
there is an additional hopping 
induced continuous diffusion which was absent when hopping was frozen. 
The time scale of relaxation 
is thus found to be faster than that predicted by hopping motion alone 
because it now also includes the contribution from the continuous dynamics.
This effect is key to showing explicitly that no strict
localization transition 
takes place at $T_{c}$, in accord with long standing arguments 
\cite{kirk,biroli}.

To keep the 
theory analytically tractable in our earlier work we 
used a simpler version of the MCT which 
included only one coupling parameter $\lambda$ and neglected the 
distribution of barrier heights in the hopping dynamics predicted by RFOT 
theory \cite{sbp}. As a result the $\alpha$ relaxation was nearly exponential. 
In this present article in 
order to address the origin of the stretching of the long time 
$\alpha$ relaxation 
dynamics we examine not only a two coupling parameter MCT 
(Gotze's $F_{12}$ model) 
having both $\lambda_{1}$ and 
$\lambda_{2}$ term which in combination can directly result 
in stretching \cite{gotze} but we also incorporate the static barrier 
height distribution, from 
RFOT theory, in the 
hopping dynamics. 
The term containing $\lambda_{1}$ 
describes the coupling of the density relaxation 
to a static field (which may describe a localized 
defect or a static inhomogeneity in the density of the
system) and that containing $\lambda_{2}$ describes the self coupling.
As shown by Gotze and co-workers, the 
$F_{12}$ model which formally describes static inhomogeneity 
present in the system predicts a stretching of the $\alpha$ 
relaxation dynamics above $T_{c}$ \cite{gotze}. 
It is not clear precisely how such a static inhomogeneity would in fact 
be generated 
above the microscopic $T_{c}$, however, such a scenario is
perhaps viable at temperatures below $T_c$ but there 
the hopping dynamics must also contribute significantly. 

 Computer simulation studies of atomic displacements in supercooled binary mixture systems 
strongly suggest 
the coexistence of continuous diffusion and hopping as mechanisms
of mass transport \cite{sarikajcp}.
These studies show that hopping events are often followed by enhanced
continuous diffusion. These two mechanisms can obviously, therefore, 
interact cooperatively with
each other. 

 The present analysis provides a quantitative description of the non-linear
interaction between continuous diffusion and dynamically heterogeneous 
activated hopping. It is shown that at and just below $T_C$,
hopping helps unlock the continuous diffusion which now becomes more effective
than the hopping would be by itself. 
We further find that below
$T_c$, the stretching of the relaxation combines the effects of activation 
inhomogeneity and static inhomogeneity. The barrier height distribution 
takes 
care of the dynamic inhomogeneity in the system and becomes the primary 
source of the stretching of 
the dynamics much below $T_{c}$, but the MCT effects play a role in
the low barrier components to enhance the rate of short time diffusion.
 
The organization of the rest of the paper is as follows. In the next section we
describe the theoretical scheme. In section III we present several analytical results
that can be derived for the combined theory. Section IV contains numerical results.
Section V concludes with a discussion on the results.


\section{Theoretical Scheme}

In our earlier article we showed that activated dynamics 
or hopping opens up an 
extra channel for the structural relaxation \cite{sbp}. The continuous dynamics was 
calculated using the one coupling parameter ($\lambda$) 
MCT.
In describing the activated 
motions the probability of a single hop was calculated from RFOT theory which 
connects the height of the free energy barrier to 
the configurational entropy. For simplicity we had considered a single value of the 
barrier height for each temperature although RFOT theory shows that this 
barrier is in fact distributed. The present theory uses the same scheme 
of calculation with some modifications to understand the relation to 
previous MCT efforts to cope with nonexponential relaxation.
$(1)$ The one coupling parameter ($\lambda$) MCT is extended to incorporate 
the two coupling parameters 
($\lambda_{1}$, $\lambda_{2}$) MCT (the $F_{12}$ model) 
which Gotze has used to address the effect of static inhomogeneity on the dynamics above $T_{c}$.
$(2)$ The single valued barrier height for the activated dynamics is 
replaced by a distribution of barrier heights in accord with RFOT theory.

The previous article used two different mathematical schemes to combine the 
hopping and the continuous dynamics (described by MCT)\cite{sbp}. 
In one of the schemes
the full intermediate scattering function was written as 
a product of a hopping 
and a MCT part using the separation of timescales between the MCT dynamics and
the hopping dynamics. In the second scheme the 
strict parallelism of hopping and continuous motion was more transparent. 
The structure of the equation in the second scheme is similar 
to that obtained by Gotze and coworkers \cite{gotze} and Das and Mazenko 
\cite{das} from more detailed microscopic derivations.
Both 
the schemes give nearly identical results. This further adds credence 
to the first scheme. In the present paper, therefore,
 we will work 
with the first scheme (easier to implement and also in this scheme 
the continuous diffusion 
and hopping dynamics can be investigated separately) 
although the extensions made here can also 
be incorporated into  the second scheme in a similar manner.

The total intermediate scattering function can be written as,
\begin{equation}
\phi(q,t)\simeq \phi_{MCT}(q,t)\phi_{hop}(q,t). \label{fqttot}
\end{equation}
\noindent
Here $\phi_{MCT}(q,t)$ is the MCT part of the intermediate 
scattering function, which is now self consistently calculated 
with $\phi(q,t)$, and its equation of motion is given by,

\begin{eqnarray}
\ddot{\bf\phi}_{MCT}(t) 
+\gamma \dot{\bf \phi}_{MCT}(t) 
&+& \Omega_{0}^{2}
{\bf\phi}_{MCT}(t)\\ \nonumber
&+& \lambda_{1} \Omega_{0}^{2} 
 \int_{0}^{t} \:dt^{\prime} {\bf\phi}(t^{\prime}) 
\dot{\bf\phi}_{MCT}(t-t^{\prime}) \\ \nonumber
&+& \lambda_{2} \Omega_{0}^{2} 
 \int_{0}^{t} \:dt^{\prime} {\bf\phi}^{2}(t^{\prime}) 
\dot{\bf\phi}_{MCT}(t-t^{\prime}) = 0 \label{fqtmct}
\end{eqnarray}
\noindent 

In the above equation the fourth term on the left hand side describes the 
coupling of $\phi_{MCT}(q,t)$ with a static field which is meant to describe
 the defects
or the inhomogeneity in the system, according to Gotze \cite{gotze}. 
The fifth term on the left describes 
the coupling of $\phi_{MCT}(q,t)$ with itself (the self coupling term).
Unlike the earlier model the present model contains two order parameters.
In the absence of hopping, 
the MCT transition would now take place not at a single point but at many points
on the ${\lambda_{1}}-{\lambda_{2}}$ plane.

In eq.\ref{fqttot} the hopping part of the intermediate scattering function 
is give by $\phi_{hop}(q,t)$. The contribution from a single hopping event 
to the scattering function was derived in our earlier paper \cite{sbp}.
It can be written as,  
\begin{eqnarray}
\phi_{hop}^{s}(q)=\frac{1}{s+K_{hop}(q)} \label{fqthop}
.
\end{eqnarray}
\noindent
where,
\begin{eqnarray}
K_{hop}(q)=\frac{P}{v_{p}}\Bigl[v_{0}-8
&&\int_{\pi/\xi}^{\infty} \:dq_{1} q_{1}^{2}
e^{-q_{1}^{2}d_{l}^{2}}\\\nonumber
&&\times \Bigl\{
\Bigl(\frac{-(q-q_{1})\xi \: cos((q-q_{1})\xi)}{(q-q_{1})^{3}}\\\nonumber
&&+\frac
{sin((q-q_{1})\xi )}
{(q-q_{1})^{3}}\Bigr)^{2}\Bigr\}\Bigr] \label{khop}
\end{eqnarray}
\noindent
In the above expression of the hopping kernel, $P$ is the average hopping 
rate which is a function of the free energy barrier height, $\Delta F$ 
and is given by $P=\frac {1}{\tau_{0}}exp(-\Delta F/k_{B}T)$ \cite{lubwoly}. 
The 
free energy barrier is calculated from RFOT theory  \cite{lubwoly}. 
$v_{0}= \frac{4}{3}\pi \xi^{3}$ 
is the region participating in hopping where $\xi$ is calculated from RFOT 
theory. $v_{p}$ is the volume of a single particle in the system. $d_{L}$ 
is the Lindemann length. In this model kernel a typical hopping event 
involves an uncorrelated 
displacement of particles by a Lindemann length. More 
complex kernels that encode correlations between movements are also possible.

Now if we consider a distribution of barrier heights then the contribution 
from multiple hoppings to the intermediate structure function can be written as,

\begin{eqnarray}
\phi_{hop}(q,t)
&=&\int \phi^{s}_{hop}(t) {\cal P}(\Delta F) d\Delta F \nonumber \\
&=&\int e^{-tK_{hop}(\Delta F)} {\cal P}(\Delta F) d\Delta F \label{hopdist}
\end{eqnarray}
\noindent

${\cal P}(\Delta F)$ is considered to be Gaussian. With a Gaussian 
distribution of barrier heights the relaxation function is known to 
fit well to a stretched exponential, where the stretching depends on 
the width of the Gaussian \cite{xiawoly}.

\begin{figure}
\epsfig{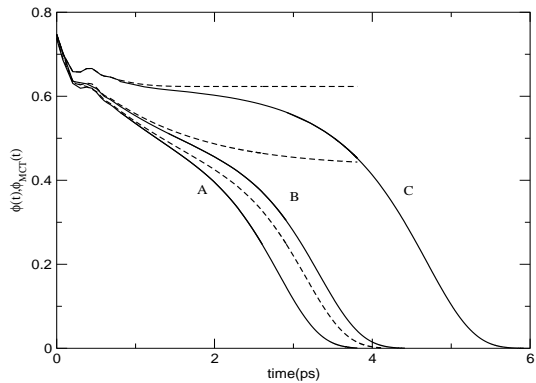}
\caption { The $\phi(t)$ calculated from the unified theory, and the results of the idealized MCT are plotted 
for three temperatures. The solid lines A, B, C correspond to the full $\phi(t)$ for 
$\epsilon=-0.1$, $\epsilon=0$ and $\epsilon=0.5$, respectively. The 
dashed lines
are the idealized MCT results. 
In this plot $\lambda=0.568 $  
which means that the MCT dynamics without hopping above $T_{c}$ is 
stretched with stretching parameter $\beta_{MCT}=0.8$. The activated 
dynamics is calculated with barrier height distribution such that 
$\beta^{static}_{hop}=0.5$.}
\end{figure}

In figure 1 we plot the results of the unified theory and also that of 
the idealized MCT ($F_{12}$ model). The plots are given for 
$\epsilon=-.1$, $\epsilon=0$ and $\epsilon=0.5$ which correspond to 
temperatures above, at and below $T_{c}$ respectively. The calculations 
are done at $\lambda=0.568$ which for idealized MCT 
above $T_{c}$ (for $\epsilon < 0$ ) 
predicts a stretched relaxation with the MCT stretching parameter, 
$\beta_{MCT}=0.8$. 
For the activated dynamics we have considered 
Gaussian distribution of barriers that would predict the stretching parameter 
for the hopping dynamics,
$\beta^{static}_{hop}=0.5$. As expected, 
hopping does not have much effect above $T_{c}$ but below $T_{c}$ the unified 
theory continues to show structural relaxation where 
the idealized MCT would have predicted  
strict localization transition. The longtime dynamics is stretched at all 
the temperatures. But as will be discussed later the stretching parameter 
for the total dynamics is different from $\beta_{MCT}$ and 
$\beta^{static}_{hop}$. Its value depends on the interaction between the 
hopping and the MCT dynamics and changes with temperature. 

\section{The effect of hopping on the MCT dynamics above, at and below $T_{c}$
:Analytical results}

In the earlier article we showed numerically that hopping has very little 
effect on the MCT dynamics above $T_{c}$ and has a nonlinear effect on the MCT 
relaxation timescale at $T_{c}$. In this article we analytically investigate
the effect of hopping on the MCT dynamics above, at and below $T_{c}$.
For this study we first examine the one parameter MCT dynamics, i.e. we 
consider $\lambda_{1}=0$. This is a special case of the two parameter model. 
The MCT transition takes place at $\lambda_{2}=4$ for $\lambda_{1}=0$. 
Initially we also take the 
barrier height distribution to be a delta function corresponding to a single 
hopping barrier. With these simplifications we use 
eqs.\ref{fqttot}-\ref{hopdist} for the following analysis of the effect of 
hopping on the MCT timescale.

When $\lambda_{1}=0$, 
Eq.\ref{fqtmct} can be rewritten as, 
\begin{eqnarray}
\frac{\Omega^{2}_{o} \phi_{MCT}(z)}{1+z\phi_{MCT}(z)}=z+i\gamma 
+ LT \lambda_{2}\Omega^{2}_{o} [\phi^{2}(t)](z)\label{fqtlap}
\end{eqnarray}
\noindent
where LT stands for Laplace transform.
Since we are interested in the $\alpha$ relaxation timescale we will take the 
longtime limit of the above equation. In the longtime limit where 
$z\rightarrow 0$, eq.\ref{fqtlap} reduces to,

\begin{eqnarray}
\frac{\Omega^{2}_{o} \phi_{MCT}(z)}{1+z\phi_{MCT}(z)}=i\gamma 
+ LT \lambda_{2}\Omega^{2}_{o} [\phi^{2}(t)](z)\label{fqtlap1}
\end{eqnarray}\noindent

In the above equation the second term on the right hand side is the self 
coupling term which makes dominant contribution at high density. But at low 
liquid density the self coupling term can be neglected and the solution of eq.\ref{fqtlap1} 
is given by, 

\begin{equation}
\phi_{MCT}(t)=e^{-K_{o}t}
\end{equation}
\noindent
where $K_{o}=\frac{\Omega^{2}_{o}}{\gamma}$ is the inverse 
timescale of longtime decay in the normal liquid regime.

Eq.\ref{fqtlap1} in the time plane can be written as,
\begin{eqnarray}
{\bf\phi}_{MCT}(t)=
\frac{1}{K_{o}}\dot{\bf \phi}_{MCT}(t) 
+\lambda_{2} 
 \int_{0}^{t} \:dt^{\prime} {\bf\phi}^{2}(t^{\prime}) 
\dot{\bf\phi}_{MCT}(t-t^{\prime}) \label{fqtalpha}
\end{eqnarray}
\noindent 
From the earlier analysis \cite{sbp,leu} we know that the solution of 
eq.\ref{fqtalpha} 
is exponential in the longtime. So we simplify $\phi_{MCT}(t)$  
using its longtime form, $\phi_{MCT}(t)=a e^{-K_{MCT}t}$, 
where $a$ is the prefactor.
Presently we also take the hopping part of the intermediate scattering function 
to be an exponential, $\phi_{hop}(t)=e^{-K_{hop}t}$. 
Using these expressions for  
$\phi_{MCT}$ and $\phi_{hop}$ in eq.\ref{fqtalpha} we get an expression 
for $K_{MCT}$ in terms of $K_{hop}, K_{o},\lambda_{2}$ and $a$.
\begin{eqnarray} 
K_{MCT}=\frac{1}{2}&\Bigl[&-(2K_{hop}-(1-a^{2}\lambda_{2})K_{o})\\\nonumber
&+&\sqrt{(2K_{hop}-(1-a^{2}\lambda_{2})K_{o})^{2}+ 8 K_{hop} K_{o}}\Bigr]\label{kmct}
\end{eqnarray}
\noindent 

In the absence of hopping the above expression reduces to,

\begin{equation} 
K_{MCT}=(1-a^{2}\lambda_{2})K_{o} \label{kmctred}
\end{equation}
\noindent
Thus we see that the self coupling term leads to an increases in the 
timescale of relaxation when compared with the bare relaxation timescale $K^{-1}_{o}$.
Eq.\ref{kmctred} further predicts that $K_{MCT}$ goes to zero or the 
relaxation time approaches infinity 
as $a^{2}\lambda_{2}$ approaches one. Thus the analysis shows that in the 
absence of hopping strict localization takes place at 
$a^{2} \lambda_{2}=1$. From previous studies we know at $T_{c}$, $a=1/2$ 
and $\lambda_{2}=4$ \cite{sbp,leu,beng}, 
thus at $T=T_{c}$, $a^{2} \lambda_{2}$ indeed becomes unity.  

We will now analyze $K_{MCT}$ in the presence of hopping 
in three different regions, above 
$T_{c}$, at and around $T_{c}$  and below $T_{c}$.

\subsection{Above $T_{c}$}

Above the mode coupling transition temperature $T_{c}$, it was shown by 
explicit calculation that the timescale of 
hopping dynamics is so much longer than direct relaxation 
in the system that it can be neglected \cite{sbp}.
The expression of $K_{MCT}$ then reduces to,
\begin{equation} 
K_{MCT}\simeq (1-a^{2}\lambda_{2})K_{o}
\end{equation}
\noindent
Thus in accord with our earlier numerical calculation \cite{sbp}  
above the mode coupling transition temperature the hopping does not have 
significant effect  on the $\alpha$ relaxation timescale.

\subsection{At and around $T_{c}$}

From the earlier studies we know that 
at the transition temperature, $a^{2}\lambda_{2}=1$ \cite{sbp,leu}. 
Thus eq.\ref{kmct} 
reduces exactly to, 

\begin{equation} 
K_{MCT}=-K_{hop}
+\sqrt{K^{2}_{hop}+ 2 K_{hop} K_{o}}
\end{equation}
\noindent
In the above expression if we take some reasonable value for $K_{o}$ and 
$K_{hop}$, we find that,

\begin{equation} 
K_{MCT}\simeq\sqrt{2 K_{hop} K_{o}} \label{kmcttc}
\end{equation}
\noindent
Thus we find that $K_{MCT}$ has a nonlinear dependence on $K_{hop}$ which is 
also in accord with our earlier numerical calculations \cite{sbp}. The coupling 
of the hopping dynamics with the short time part of the MCT dynamics 
(liquid like dynamics) leads to this nonlinear
dependence. Thus we find that coupling between the short time part of the 
MCT dynamics and hoppings leads 
to an MCT part of the structural relaxation timescale 
which is much faster than the hopping 
timescale. This is a critical effect and is found at and near $T_{c}$.  
This result corroborates our earlier findings that a 
single hopping event leads to many continuous 
diffusion events which are the primary means of structural relaxation 
in this region.
In the next subsection we will find that this scenario changes as we go lower 
and lower in temperature.

\subsection{Below $T_{c}$}

Much below the transition temperature, $a^{2}\lambda_{2}>>1$. 
Eq.\ref{kmct} can be rewritten as, 
\begin{eqnarray} 
K_{MCT}=\frac{1}{2}&\Bigl[&((a^{2}\lambda_{2}-1)K_{o}+2K_{hop})\\\nonumber
&&\times{\Bigl\{} -1+
\sqrt{1+\frac {8 K_{hop} K_{o}}{((a^{2}\lambda_{2}-1)K_{o}+2K_{hop})^{2}}}
{\Bigr\}}
\Bigr]\label{kmct1}
\end{eqnarray}
\noindent
Since at low temperatures $((a^{2}\lambda_{2}-1)K_{o}+2K_{hop})^{2} 
>> 8 K_{hop} K_{o}$ thus we can write, 
\begin{equation}
K_{MCT}=\frac {2 K_{hop} K_{o}}{((a^{2}\lambda_{2}-1)K_{o}+2K_{hop})}
\end{equation}
\noindent
Now if we further take into consideration that $(a^{2}\lambda_{2}-1)K_{o} >> 
2K_{hop}$ then the above equation reduces to,
\begin{equation} 
K_{MCT}\simeq \frac{2 K_{hop}}{(a^{2}\lambda_{2} -1)}\label{kmctbelow}
\end{equation}
\noindent
Thus much below the transition temperature the timescale of the MCT dynamics 
and also the total dynamics becomes slaved to the hopping timescale.
Analyzing eq.\ref{kmctbelow}, an important observation can be made about 
the relaxation timescale. It is known that as we lower the temperature both 
$\lambda_{2}$ and $a$ increases. Thus the denominator in 
eq.\ref{kmctbelow} will increase as we lower the temperature. From the 
analysis in the earlier subsection we know that initially to 
start with, below $T_{c}$ $K_{MCT}>> K_{hop}$, then as we keep lowering 
the temperature then depending on the temperature dependence of $\lambda_{2}$ 
and $a$ ,$K_{MCT}\simeq K_{hop}$. But as we further lower the temperature then 
slowly $K_{MCT}<< K_{hop}$. Thus in this regime although there will be hopping 
induced continuous diffusion but the primary mode of the structural relaxation 
becomes direct activated hopping itself. 

\section {Interplay between hopping and MCT dynamics at and below $T=T_{c}$ 
: Numerical results}

Continuing from our analysis where both MCT and hopping dynamics are assumed 
to be exponential, 
here we will 
present some numerical results for the general case where both hopping and 
MCT dynamics can be stretched. For these general cases we will try to 
understand the effect of hopping on the MCT dynamics. 

For the calculation we need to solve  eq.\ref{fqtmct} numerically. 
It is well known that due to the self consistent nature of the equation, 
its numerical solution becomes a nearly Herculean task around 
the mode coupling 
transition temperature. In the presence of hopping due to the disparate 
timescales present in the system and the convolution in eq.\ref{fqtmct} 
which involves all these timescales, the time of calculation increases 
many fold. However the scheme proposed by Fuchs {\it et al}
\cite{hofac} allows the calculation to be done much faster. 
Both $\phi(t)$ 
and $\phi_{MCT}(t)$ vary more slowly for longer times than they do 
at short times.
The essential idea involved in the scheme is to separate the slow and the fast 
variables and treat them differently in the convolution. 
The short time part of $\phi(t)$ and $\phi_{MCT}(t)$ are calculated exactly
with very small stepsize and they are then used as input to carryout 
the calculation for the long time part of the same. We have exactly 
followed the scheme presented 
in reference \cite{hofac} except for a minor modification (one extra term)
in eq.29 of 
reference \cite{hofac} when the integration timestep is an odd number.
In our calculations the total timestep N=1000.

For this study we have 
picked three points on the $\lambda_{1}$, $\lambda_{2}$ plane. 
In all the cases 
to understand the temperature effect the values 
of $\lambda_{1}$, $\lambda_{2}$  were varied according to the expressions given
bellow,

\begin{eqnarray}
\lambda_{1}= \frac {(2 \lambda -1)}{\lambda^{2}}+ \epsilon \frac {\lambda}
{(1+(1-\lambda)^{2})} \label{l1}
\end{eqnarray}

\begin{eqnarray}
\lambda_{2}= \frac {1}{\lambda^{2}}+ \epsilon \frac {\lambda (1-\lambda)} 
{(1+(1-\lambda)^{2})}\label {l2}
\end{eqnarray}
\noindent
where the value of $\lambda$ determines the value of  
$\lambda_{1}$ and $\lambda_{2}$ at $T=T_{c}$. $\epsilon$ is a measure of 
the distance from the MCT transition temperature. As in our earlier model 
calculation, \cite{sbp} the values of $\Omega^{\star}_{o}$ and $\gamma^{\star}$ 
are kept unity and the scaling time as 1ps. In the numerical calculations,
to understand the temperature effect although we have changed the 
values of $\lambda_{1}$ and $\lambda_{2}$ but we have kept all the other 
parameters constant including the hopping barriers. 
For the calculation of the hopping part we have used eq.\ref{hopdist}
with a Gaussian distribution of barriers. 
According to the RFOT theory, the mean barrier height and thus the hopping 
timescale should change 
with temperature \cite{xiawoly,lubwoly}.
In its simplest form without taking barrier softening into consideration the 
mean barrier height can be written as, $\Delta F/k_{B}T=32 k_{B}/s_{c}=
\frac {32 k_{B}}{\Delta c_{p}} \frac {T_{K}}{(T-T_{K})}$, where $s_{c}$ is the 
configurational entropy, $\Delta c_{p}$ is the jump in the specific heat and 
$T_{K}$ is the Kauzmann temperature \cite{xiawoly,lubwoly}. However in 
this present calculation to clearly assign any change in dynamics due to 
the change in $\lambda_{1}$ and $\lambda_{2}$ value we have kept the 
mean of the distribution  
fixed at about 8.8 $k_{B}T$.
As mentioned earlier the stretching in the hopping dynamics 
is determined by the width of the Gaussian distribution of the barriers. The 
broader the distribution the more stretched is the dynamics. We have varied 
the width of the distribution such that the $\beta^{static}_{hop}$ varies 
from 0.2 to 0.8. Along with the $\beta^{static}_{hop}$ value the timescale of hopping
also changes, which has been taken into account in our calculation.

For all the cases the $\beta_{total}$ values are plotted against $\beta^{static}_{hop}$
for different $\epsilon$ values, where $\beta_{total}$ is the stretching 
parameter for $\phi(t)$ and $\beta^{static}_{hop}$ is the same for $\phi_{hop}(t)$. 

\subsection {Case 1: $\lambda=0.5$} 

In the first case we consider an example when $\lambda=0.5$. 
This value of $\lambda$ 
implies that at $T=T_{c}$, $\lambda_{1}=0$ and $\lambda_{2}=4$ and the 
MCT  dynamics just above $T_{c}$ is exponential. For this case we vary   
$\epsilon$ from $0-1$. 
In figure 2, the $\beta_{total}$ values are plotted against $\beta^{static}_{hop}$
for the three $\epsilon$ values. As expected for $\epsilon=0$ the 
$\beta_{total} \simeq 1$. This is because although without hopping structural 
relaxation is arrested but once hopping is present the relaxation is  
dominated by the MCT dynamics. As we increase the epsilon value we find that 
the effect of hopping is stronger and the dynamics gets stretched. But we also 
notice that for very low $\beta^{static}_{hop}$ value the dynamics is less stretched 
than the static barrier distribution would indicate.
This is because lower $\beta^{static}_{hop}$ value implies a broader 
barrier height distribution which means we populate both lower and 
higher barrier heights. 
The small barrier hoppings actually 
couple to the liquid like part of the MCT dynamics (whose timescale 
is given by $K_{o}^{-1}$) and the relaxation is 
faster and dominated by 
this liquid like MCT dynamics which has a much shorter timescale. 
As we increase the $\epsilon$ value MCT gets more slaved to hopping and thus 
the time scale of relaxation increases and the liquid like MCT dynamics becomes
less important. If we further lower the temperature (or increase the $\epsilon$
value) we would find that even for small $\beta^{static}_{hop}$ values the total 
dynamics is determined primarily by hopping.

\begin{figure}
\epsfig{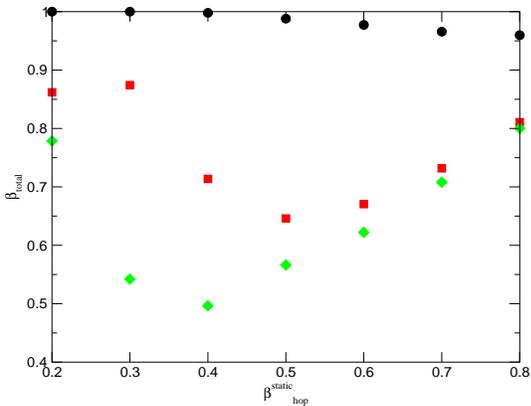}
\caption{The stretching parameter for the total structural relaxation, 
$\phi(t)$, $\beta_{total}$ is plotted against the same for $\phi_{hop}(t)$,
$\beta^{static}_{hop}$ for three different $\epsilon$ values. The solid circle is for 
$\epsilon=0$, the solid square is for $\epsilon=0.5$ and the solid diamond 
is for $\epsilon=1$. In this plot the $\lambda=0.5 $  
which means that the MCT dynamics without hopping above $T_{c}$ is 
exponential.} 
\end{figure}
The interplay of hopping with MCT nonlinearities can be thought of as 
a quantitative formulation of "facilitation effects" \cite{xiawoly}. 
As Xia and Wolynes pointed out hopping events interact if they occur 
near each other. This is accounted for by the MCT nonlinearity. 
In the Xia-Wolynes treatment the corresponding effect led to the 
cutoff of the relaxation time distribution, on the slow side, 
owing to the renewal of a mosaic cell's environment through hops. 
This resulted in an increased $\beta$ from that obtained from 
the static Gaussian model, 
as occurs here too \cite{xiawoly}.

\begin{figure}
\epsfig{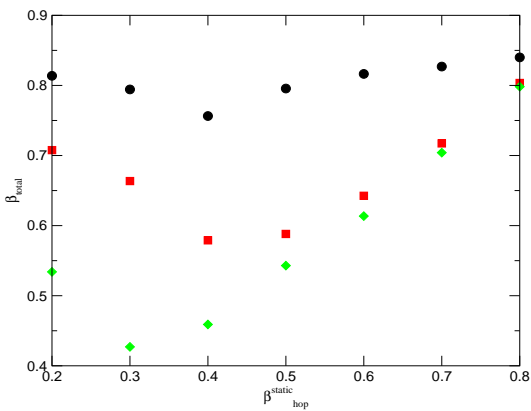}
\caption{The stretching parameter for the total structural relaxation, 
$\phi(t)$, $\beta_{total}$ is plotted against the same for $\phi_{hop}(t)$,
$\beta^{static}_{hop}$ for three different $\epsilon$ values. The solid circle is for 
$\epsilon=0$, the solid square is for $\epsilon=0.5$ and the solid diamond 
is for $\epsilon=1$. In this plot the $\lambda=0.568 $  
which means that the MCT dynamics without hopping above $T_{c}$ is stretched 
with stretching parameter, $\beta_{MCT}=0.8$. }
\end{figure}

\subsection {Case 2: $\lambda=0.568$} 

In this case we consider that $\lambda=0.568$. This value of $\lambda$ 
implies that at $T=T_{c}$, $\lambda_{1}=.4215$ and $\lambda_{2}=3.09958$ 
and the MCT dynamics, (without hopping) just above $T_{c}$, would already 
be stretched 
with $\beta_{MCT}=0.8$. For this case we vary  
$\epsilon$ from $0-1$. 
In figure 3, the $\beta_{total}$ values are plotted against $\beta^{static}_{hop}$
for the three $\epsilon$ values. The results are similar to those obtained 
for the first case. For $\epsilon=0$ the 
$\beta_{total} \simeq 0.8 $ which implies that the total dynamics is 
still determined primarily by the MCT dynamics. We also find that for 
high $\beta^{static}_{hop}$ values, as $\epsilon$ increases, 
the dynamics gets more 
and more dominated by the hopping dynamics. Nevertheless, 
for small $\beta^{static}_{hop}$ 
the scenario is a little different. For smaller $\epsilon$ values 
the low barrier hoppings couple to the liquid like part of the MCT dynamics
and the  dynamics is less stretched and also much faster. However,
as we increase 
$\epsilon$, MCT dynamics gets more and more slaved to hopping dynamics and the 
MCT relaxation timescale becomes proportional to the hopping relaxation 
timescale
(similar to that shown in Eq.\ref{kmctbelow}). Thus the effect of the coupling 
of low barrier hoppings with the liquid like part of the MCT dynamics, on the 
total MCT dynamics, reduces. If we further lower the temperature then we would find 
that even for $\beta^{static}_{hop}=0.2$ the total dynamics follows the hopping dynamics.

The results in case 1 and case 2 look quite similar but more
detailed observation reveals that the value of $\beta^{static}_{hop}$, 
where $\beta_{total}$ 
begins to increase, is smaller for case 2 (where MCT dynamics itself is 
more stretched) than it is in case 1. As discussed earlier, the reason 
$\beta_{total}$ increases for small $\beta^{static}_{hop}$ is that the 
small barrier 
hopping gets coupled to the liquid like part of the MCT dynamics 
allowing the 
total structure to relax. Now in case 2, the MCT dynamics is 
itself stretched thus 
the effect of the small barrier hopping on the MCT dynamics will be much less
effective when compared to case 1. This trend becomes clearer when we 
study the next case where the MCT dynamics is much more stretched.

\subsection {Case 3: $\lambda=0.75$} 

In this case we consider that $\lambda=0.75$. This value of $\lambda$ 
implies that at $T=T_{c}$, $\lambda_{1}=.889$ and $\lambda_{2}=1.778$ 
and the MCT dynamics (without hopping) just above $T_{c}$ would already 
be  stretched  
with $\beta_{MCT}=0.5$. For this case we vary  
$\epsilon$ from $0-1$. 
In figure 4, the $\beta_{total}$ values are plotted against $\beta^{static}_{hop}$
for the three $\epsilon$ values. The results are similar to that obtained 
for case 1 and case 2 for $\epsilon=0$. But for higher $\epsilon$ values unlike 
in case 1 and 2,
$\beta_{total}$ continuously decreases with $\beta^{static}_{hop}$. This is because 
as discussed before, since MCT dynamics is already stretched, the structural 
relaxation due to low barrier hopping is less effective.
 Also note that 
the $\beta_{total}$ decreases with $\beta^{static}_{hop}$ value but it is 
neither equal to $\beta^{static}_{hop}$, nor equal to the pre-transition 
$\beta_{MCT}$ value. At these temperatures although MCT dynamics is slaved to hopping 
but both the channels of relaxation are almost equally effective. If we 
compare the $\beta_{total}$ values for $\epsilon=$ 0.5 and 1, we will find that 
in most of the cases $\beta_{total}$ for $\epsilon=1$ has a lower value.
This is because at lower temperatures MCT dynamics becomes a less effective 
relaxation channel.
At further lower temperatures (higher $\epsilon$ values) $\beta_{total}$ will 
follow $\beta^{static}_{hop}$ more closely.

\begin{figure}
\epsfig{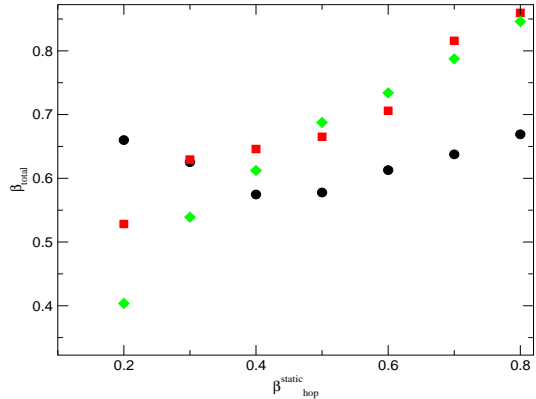}
\caption{The stretching parameter for the total structural relaxation, 
$\phi(t)$, $\beta_{total}$ is plotted against the same for $\phi_{hop}(t)$,
$\beta^{static}_{hop}$ for three different $\epsilon$ values. The solid circle is for 
$\epsilon=0$, the solid square is for $\epsilon=0.5$ and the solid diamond 
is for $\epsilon=1$. In this plot the $\lambda=0.75 $  
which means that the MCT dynamics without hopping above $T_{c}$ is stretched 
with stretching parameter, $\beta_{MCT}=0.5$.} 
\end{figure}
\section{Concluding Remarks}
 
MCT and the RFOT theory provide a unified 
theory of relaxation over the whole temperature plane \cite{sbp}. Even 
without dynamical heterogeneity the theory 
successfully predicts the decay of the structural relaxation below mode 
coupling transition temperature, $T_{c}$,  
confirming there is no strict localization transition at $T_{c}$. 
Without dynamical heterogeneity of the instantons the coupled theory lead
to an exponential $\alpha$ relaxation, but in the laboratory 
the $\alpha$ relaxation is generally stretched. 
In the present article we examined both a two parameter 
MCT model \cite{gotze} and more realistically one that also included 
a barrier height distribution that gives rise to stretching in the 
hopping dynamics by itself \cite{xiawoly}.

The study has been carried out for different stretching parameters of the MCT 
dynamics, that is by changing $\lambda$ values in the $F_{12}$ model 
and also for different stretching parameters of the hopping dynamics obtained 
by changing the width of the distribution of barrier heights.

 To summarize, the main conclusions of the present work is that the
continuous dynamics, described here within the $F_{12}-$MCT formalism
, and the
activated hopping dynamics, described here using RFOT theory, interact
in a non-linear fashion to give rise to dynamical features which are 
distinct from both. MCT by itself of course cannot describe dynamics
below its
critical temperature. We find that hopping facilitates the continuous dynamics
channel and in the process the effects of hopping on the relaxation
decreases. Thus, one finds the stretching parameter arising 
solely from distribution of hopping barrier energies in RFOT is increased 
by the mode coupling terms.  
It is also found that when MCT dynamics is less stretched then the 
effect of hopping
on the MCT dynamics is less pronounced and the hopping dominated regime 
moves to a lower temperature. On the other hand,
 for more stretched MCT dynamics, due to 
the larger overlap of MCT and hopping timescales, hopping begins to dominate 
at a higher temperature.

 We have already mentioned the need for using a more complex hopping kernel
than that given by Eqs.3-5. In particular, one needs to include the effects of
mode coupling softening on the barrier height distribution. 
This is a feed-back effect of unleashing the mode coupling 
relaxation channels due to hopping, on the barrier height 
distribution itself. This non-linear feed-back
is expected to shift the distribution to lower barrier heights and in turn
accelerate mode coupling relaxation which can further enhance hopping. The whole
system of equations needs to be solved self-consistently. To achieve this,
we need to understand more quantitatively the effects of softening 
on the barrier height distribution.

{\bf ACKNOWLEDGEMENT}

 This work was supported in parts from NSF (USA) and DST (India).


\begin{references}

\bibitem{sbp} S. M. Bhattacharyya, B. Bagchi, P. G. Wolynes, Phys. Rev. E 
{\bf 72} 031509 (2005).

\bibitem{gotze}  W. Gotze and L. Sjogren, Z. Phys. B- Cond. Mat., {\bf 65}, 
415 (1987);W. Gotze and L. Sjogren, J. Phys. C:Solid State Phys. {\bf 21}, 3407 (1988). 


\bibitem{das}S. P. Das and G. F. Mazenko, Phys. Rev. A, {\bf 34}, 2265 (1986).


\bibitem{kirk} T.R. Kirkpatrick and P.G. Wolynes, Phys. Rev. {\bf E 35}, 
3072 (1987).

\bibitem{biroli} J. P. Bouchaud and G. Biroli, J. Chem. Phys. {\bf 121},
7347 (2004).


\bibitem {sarikajcp}S. Bhattacharyya, A. Mukherjee, and B. Bagchi, 
J. Chem. Phys.{\bf 117}, 2741(2002); S. Bhattacharyya and B. Bagchi, Phys. Rev. Lett. {\bf 89}, 
025504-1(2002); A. Mukherjee, S. Bhattacharyya and B. Bagchi, J. Chem. Phys. 
{\bf 116}, 4577 (2002); 


\bibitem{xiawoly} X. Xia and P. G. Wolynes, Phys. Rev. Lett. {\bf 86}, 5526 
(2001);Proc. Natl. Acad. Sci. U. S. A. {\bf 97}, 2990 (2000).


\bibitem{lubwoly}V. Lubchenko and P. G. Wolynes, J. Chem. Phys. {\bf 119}, 
9088 (2003);

\bibitem{leu} E. Leutheusser, Phys. Rev. A {\bf 29}, 2765 (1984).

\bibitem{beng} U. Bengtzelius, W. Gotze, and A. Sjolander, J. Phy. C {\bf 17},
5915 (1984).

\bibitem{hofac}M. Fuchs, W. Gotze, I. Hofacker and A.Latz, J. Phys: Condensed Matter {\bf 3}, 5047 (1991).

\end{references}
\end{document}